\documentclass[letter]{aa} 

\usepackage{graphicx}
\usepackage{txfonts}
\usepackage{hyperref}
\addto\extrasenglish{
    
}
\addto\extrasenglish{
    
}
\hypersetup{
    colorlinks = true,
    linkcolor = {purple},
    citecolor = {blue}
}
\usepackage{graphicx}
\usepackage{txfonts}
\newcommand{\gppr}{\stackrel{>}{\scriptstyle \sim}}
\newcommand{\gappr}{\raisebox{-0.4ex}{$\gppr$}}

\newcommand{\mesa}{MESA}

\newcommand{\cv}{CV}
\newcommand{\cvs}{CVs}

\newcommand{\msun}{M$_\odot$}

\defcitealias{rappaportetal83-1}{RVJ}

\begin{document} 

\definecolor{purple}{rgb}{0.4, 0.0, 0.8}

   \title{A modest change in magnetic braking at the fully convective boundary explains cataclysmic variable evolution}
   \titlerunning{Saturated magnetic braking and CV evolution revisited}

   \author{Joaquín A. Barraza-Jorquera
          \inst{1,2}\fnmsep\thanks{Corresponding author; joaquin.barrazaj17@gmail.com},
          Matthias R. Schreiber \inst{1}, 
          Stuart Littlefair \inst{3},
          Diogo Belloni \inst{4}
          \and
          Axel D. Schwope \inst{5}
          }
    \authorrunning{Barraza-Jorquera \& Schreiber et al. }

   \institute{Universidad T\'ecnica Federico Santa Mar\'ia (UTFSM), Department of Physics, Valpara\'iso, Chile
   \and
  Pontificia Universidad Católica de Valparaíso (PUCV), Institute of Physics, Valparaíso, Chile
  \and
  Department of Physics and Astronomy, University of Sheffield, Sheffield S3 7RH, UK
  \and
  International Centre of Supernovae (ICESUN), Yunnan Key Laboratory of Supernova Research, Yunnan Observatories, CAS, Kunming 650216, China
        \and
        Leibniz-Institut für Astrophysik Potsdam (AIP), An der Sternwarte 16, 14482, Potsdam, Germany
        }

   \date{Received XX, XXXX; accepted XX, XXXX}
 
  \abstract
   {For decades, reproducing the orbital period distribution of non-magnetic Cataclysmic Variables (CVs) seemed to require a drastic decrease, usually termed disruption, of angular momentum loss through magnetic braking at the fully convective boundary, which argued for a change in the dynamo mechanism operating in fully and partially convective stars. However, recent studies showed that the magnetic braking prescription traditionally used in CV evolution theory is clearly outdated as saturation, that is, a weak period dependence for rapidly rotating stars, is not included. }
   {Here we test an updated version of a saturated magnetic braking prescription that has been developed to explain the spin-down of single stars in the context of CV evolution. This prescription contains a boosting and a disruption parameter that represent the change in the strength of magnetic braking at the fully convective boundary.}
   {We performed state of the art \mesa~simulations for CVs with the revised saturated magnetic braking prescription. }
   {As in previous studies, we found that magnetic braking needs to be stronger in close binaries than in single stars and that, in contrast to what is observed in single stars, magnetic braking needs to be reduced at the fully convective boundary. However, in contrast to previous studies of CV evolution, only a moderate disruption by a factor of $2-3$ is sufficient to explain key features of the CV orbital period distribution and the measured mass-radius relation for CV donors. }
   {The relatively small decrease of the efficiency of magnetic braking at the fully convective boundary might have implications for our understanding of dynamo models for fully and partially convective stars. 
   }
   \maketitle

\section{Introduction}

For decades, a central assumption in the standard model of close white dwarf binary evolution has been a sharp and drastic reduction in angular momentum loss at the fully convective boundary.
This assumption is mostly based on the observational fact that only few cataclysmic variables (CVs), 
close binary systems in which a white dwarf accretes from a Roche-lobe filling low-mass companion star, have been found in the period range between $\sim2$ and $\sim3$ hours, a feature that has been termed the orbital period gap 
\citep[e.g.][]{schreiberetal24-1}. 

The orbital period gap can be explained by assuming strong angular momentum loss, which drives high mass-transfer rates in donor stars that still possess a radiative core, followed by a drastic reduction in angular momentum loss once the donor becomes fully convective. The initially high mass-transfer rates drive the donor star out of thermal equilibrium and its radius becomes inflated relative to that of single stars of the same mass. When angular momentum loss decreases sharply, the donor reestablishes thermal equilibrium, contracts, and detaches from its Roche lobe. The resulting detached system then evolves through the period gap driven by angular momentum loss through gravitational radiation without undergoing further mass transfer
until at an orbital period of $\sim2$\,hr the Roche lobe has sufficiently shrunk to trigger Roche lobe overflow \citep[e.g.][]{belloni+schreiber23-1}. 

To the best of our knowledge, the only mechanism that can produce strong angular momentum loss for donor stars with a radiative core is magnetic braking.
To reproduce the orbital period gap of CVs it is therefore usually assumed that magnetic braking completely ceases or is at least drastically reduced 
at the fully convective boundary 
\citep[e.g.][]{rappaportetal83-1}. 

Assuming that magnetic braking is responsible for the high mass transfer rates and also the reduction of angular momentum loss at the fully convective boundary, also nicely explains why the period gap is much less pronounced, or maybe even absent, in CVs that host a strongly magnetic white dwarf 
\citep{schreiberetal21-1,schreiberetal24-1,schwope25-1}
as the strong white dwarf magnetic field can reduce the wind zones of the donor star
\citep{webbink+wickramasinghe02-1,bellonietal20-1}.
Furthermore, it seems that CVs indeed evolve through the orbital period gap as detached systems \citep{zorotovicetal16-1} and that in close detached binaries consisting of a white dwarf with an M dwarf companion angular momentum loss is indeed much lower if the companion is fully convective \citep{schreiberetal10-1,shariat+el-badry26-1}.

However, there is growing evidence that the prescriptions for magnetic braking used in the standard scenario of CV evolution is not adequate. First of all, \citet{elbadryetal22-1} showed that the period distribution of close main sequence binary stars is flat which can only be explained if magnetic braking saturates, that is, if the period dependence of magnetic braking for close binaries is much shallower than assumed in the prescription proposed by \citet{rappaportetal83-1}. 
Such a saturation has been previously observed in various magnetic activity indicators and across the fully convective boundary \citep[e.g.,][]{reinersetal09-1,wrightetal11,maguaddaetal20-1,medinaetal20-1}.
Even worse, in contrast to the most fundamental assumption in the standard scenario for CV evolution, in single M dwarfs angular momentum loss seems to increase by a factor of $\sim1.5$\,across the fully convective boundary \citep{luetal24-1, chitietal24-1}.

Recently, CV evolution theory has started to take into account saturation. In particular, 
\citet{bellonietal24-1} showed that a boosted and disrupted but saturated prescription can explain observations of detached white dwarf binaries 
and  
\citet{barrazaetal25} showed that the very same prescription works for CVs. 
While including saturation clearly represent a step forward towards a unified magnetic braking prescription, saturated prescriptions used in close binaries still require a dramatic decrease (by a factor $30-100$) of the efficiency of magnetic braking, the so-called disruption, at the fully convective boundary \citep{bellonietal24-1,blombergetal24-1,barrazaetal25}.

The saturated magnetic braking prescription used in close binaries so far are all based on the early work by \citet{sillsetal00-1}. Alternative saturated models discussed for single stars have so far been largely ignored.  
We here present results from implementing an improved version of the saturated magnetic braking model suggested by \citet{mattetal15-1} for single stars in the context of CVs. 
While both models consider saturation of magnetic braking, the prescription from \citet{mattetal15-1} has a stronger dependence on the mass and radius of the star 
and no dependence on the convective turnover time in the saturated regime.

\section{Recalibrating saturated magnetic braking}

The most important feature for all saturated magnetic braking prescriptions is the critical rotation period ($P_{\rm crit}$) for which saturation occurs. 
Using semi-empirical relations, \citet{wrightetal11}
demonstrated that $P_{\rm crit}$ depends on the convective turnover time as $P_{\rm crit} = R_{\rm o , crit} \times \tau_{\rm c}$ and estimated the critical Rossby number to be $R_{\rm o , crit}\approx 0.13$.
This calibration has been widely used in different contexts. However, recently it was shown that the used semi-empirical convective turnover time significantly differs from theoretical values obtained 
from mixing length theory \citep{landinetal23-1,gossageetal25-1}.

\begin{figure*}
    \centering
    \includegraphics[width=\linewidth]{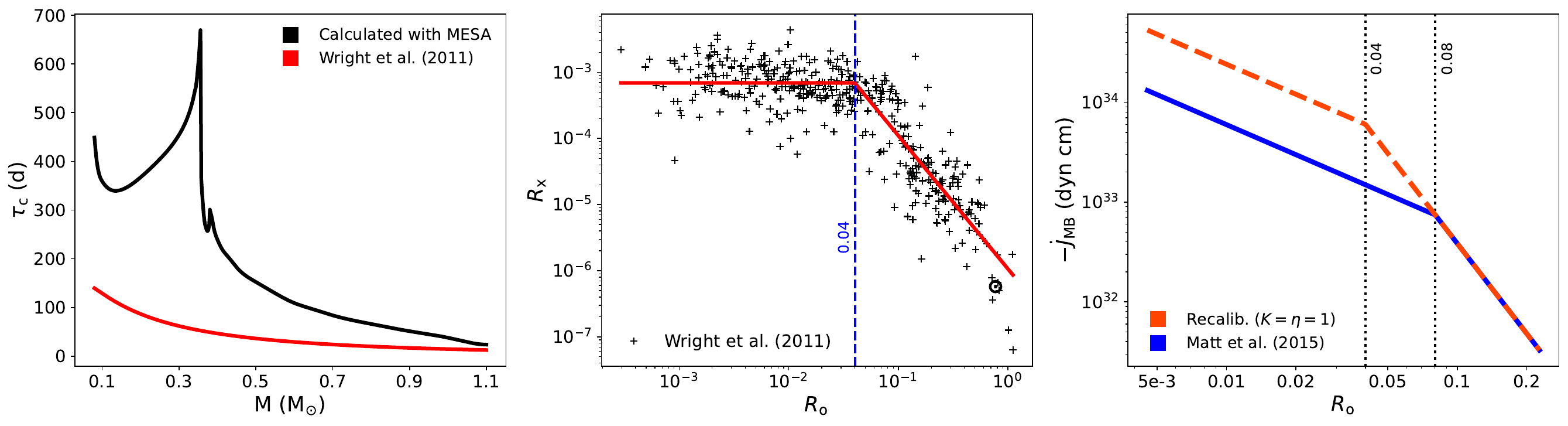}
    \caption{A new saturation threshold appears naturally from fitting observational data using calculated turnover times. The global convective turnover time is very different to frequently used approximations \citep{wrightetal11}, especially around and below the fully convective boundary (left panel). A lower threshold ($0.04$) for the Rossby number separating the saturated from the non-saturated regime is derived from fitting observations using the calculated global convective turnover time (middle panel). 
    This causes magnetic braking to be stronger in the saturated regime as we keep the strength and slope of magnetic braking in the unsaturated regime
    unchanged (right panel).}
    \label{fig:saturation}
\end{figure*}

In this work we therefore calculate the global convective turnover time directly in MESA at each step of the evolution following the equation defined by \citet{kim+demarque96}:
\begin{equation}
    \tau_{{\rm c}} = \int_{R_{\rm b}}^{R_*} \frac{{\rm d}r}{v_{\rm c}} ,
    \label{eq:tauc}
\end{equation}
where $R_{\rm b}$ and $R_*$ are the radius of the base of the convective envelope and the radius of the surface of the star respectively, and $v_{\rm c}$ is the convective velocity. We calculated $\tau_{\rm c}$ for the same set of stars used in \citet{wrightetal11} to determine the critical Rossby number. For each mass we used the \mesa~code to calculate the age at the zero-age main-sequence defined as the moment when $L_{\rm nuc}/L_{\rm tot} > 0.99$ \citep[see][]{Paxton2011}. For simplicity, we excluded the stars that would still be on the pre-main sequence according to our calculations and for field stars we assumed an age of $1$ Gyr (the exact value of the assumed age does not affect our results).
The saturation threshold we obtain this way is much smaller, that is, $R_{\rm o , crit} \approx 0.04$ (see Fig. \ref{fig:saturation}). This result is consistent with the results from \citet{argiroffietal16}. 

Saturated magnetic braking models have traditionally been calibrated for a critical rotation period defined as $P_{\rm crit} = 0.1 \, P_{\odot} \, \tau_{\rm c}/\tau_{\odot}$, where the factor $0.1$ comes from the value of $R_{\rm o, crit}$ estimated by \citet{wrightetal11}. The values $P_{\odot}$ and $\tau_{\odot}$ are included to reproduce the observed rotation period of the Sun. Calibrating magnetic braking to match the solar rotation rate 
can introduce significant inconsistencies because there is no consensus of the real value of $\tau_\odot$ 
\citep{jaoetal22-1}. Therefore, we adopt a different equation for the critical rotation period including the revised value of the critical saturation threshold and with no dependence on $\tau_\odot$ and $P_{\odot}$, that is, 
$P_{\rm crit} \ = \ 
0.04 \times \tau_{\rm c}.$

This threshold is combined with the saturated magnetic braking prescription suggested by \citet{mattetal15-1}, which is based on the general equation of the torque derived from models of stellar wind dynamics \citep{kawaler88,mattetal12}, the effects of magnetic field geometry, and the wind acceleration profile \citep{revilleetal15}. The resulting magnetic braking law is: 
\begin{equation}
\dot{J}_{\rm SAT} \, = 
{\rm T_0}
\left( \frac{R_*}{{\rm R}_{\odot}} \right)^{3.1}
\left( \frac{M_*}{{\rm M}_{\odot}} \right)^{0.5}
\left( \frac{\tau_{{\rm c}}}{\rm d} \right)^{2}
\left\{
\begin{array}{ll}
P_{\rm rot}^{-3},                     & {\rm unsaturated}, \\
P_{\rm rot}^{-1}\,P_{\rm crit}^{-2},  & {\rm saturated},
\end{array}
\right.
\label{eq:MBsat}
\end{equation}
where ${{\rm T_0} = - 1.135\times10^{32}} \, \rm{erg}$ is a calibrated constant assuming $P_{\rm crit}$ as defined above
and our estimated value of $\tau_\odot = 34.869$ days. $M_*$, $R_*$, and $P_{\rm rot}$ are the mass, radius, and rotation period (in days) of the companion main-sequence star, respectively, and $\tau_{\rm c}$ is the global convective turnover time of the companion star. 

Finally, we add the previously introduced boosting (for stars with a radiative core) and disruption (for fully convective stars) factors $K$ and $\eta$ \citep{bellonietal24-1,barrazaetal25}:
\begin{equation}
\dot{J}_{\rm SBD} \, = \,
\left\{
\begin{array}{ll}
K \, \dot{J}_{\rm SAT} \, ,         & {\rm (partially~convective)}, \\
 & \\
\left(K \, \dot{J}_{\rm SAT}\right)/\eta \, , & {\rm (fully~convective}).
\end{array}
\right.
\label{eq:MBrecipe}
\end{equation}
This prescription defines our revised saturated,
boosted, and disrupted (hereafter SBD) magnetic braking law.

\section{CV evolution with the revised SBD model}

We used the Modules for Experiments in Stellar Astrophysics (\mesa) code \citep[][version 24.03.1]{
Jermyn2023} to compute the evolution of \cvs~to test the prescription of magnetic braking suggested by \citet{mattetal15-1} with the modifications of boosting and disruption \citep{barrazaetal25}. 
We assumed solar metallicity for the donor star ($Z_{\odot} = 0.02$) and standard assumptions for \cv~evolution. 
Roche lobe radii in binary systems are computed using the fit of
\citet{Eggleton1983}. Mass transfer rates through Roche lobe
overflow in binary systems were determined following the
prescription of \citet{Ritter1988}.
Systemic angular momentum loss through the emission of gravitational waves was included as described in \citet[][]{Paczynski1967}. 
We assumed that the white dwarf is a point mass and its mass remains constant at a typical CV mass of WD $M_{\rm WD} = 0.83$ \msun~\citep{zorotovicetal11-1}, that is, that the same amount of mass that is accreted during a nova cycle is expelled during the eruption in rough agreement with model predictions \citep[e.g.,][]{yaronetal05-1}.

Additionally, it has been shown that consequential angular momentum loss (CAML) can cause \cvs~with low-mass white dwarfs to evolve into dynamically unstable mass transfer, which brings the predicted and observed white dwarf mass distributions of \cvs~into agreement and leads to a substantially reduced predicted space density \citep{schreiberetal16-1, bellonietal18-1}.
We therefore include angular momentum loss according to the empirical relation for CAML proposed by \citet{schreiberetal16-1}. 
Our \mesa\,inlists are made available at Zenodo\footnote{https://doi.org/10.5281/zenodo.19010875}. 
\begin{figure*}
    \centering
    \includegraphics[width=\textwidth]{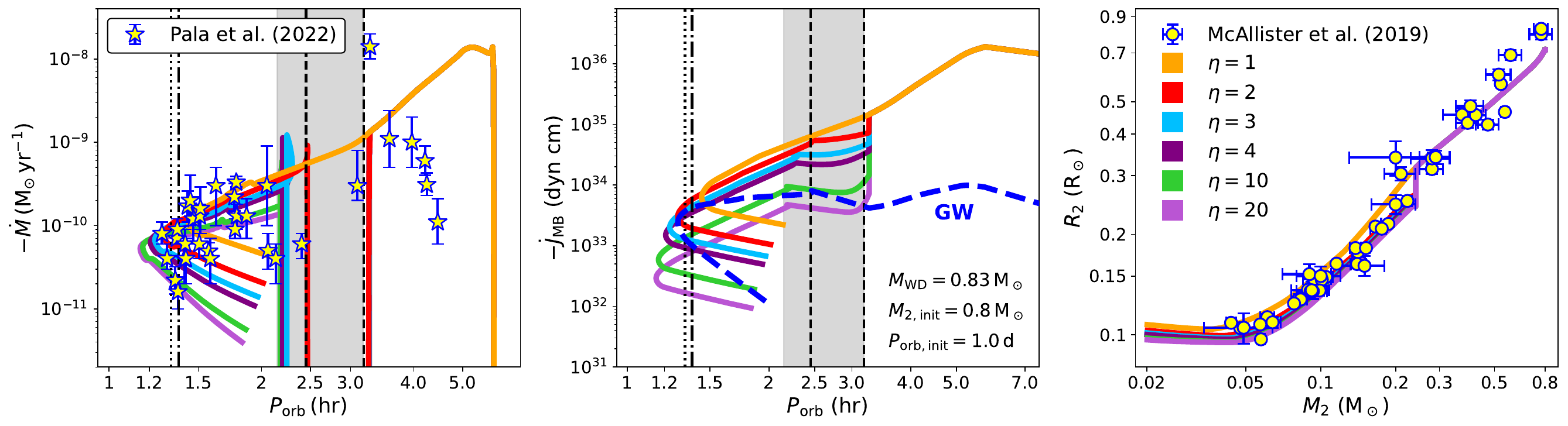}
    \caption{Mass transfer rates (left) and angular momentum loss rates (middle) as a function of orbital period as well as the mass-radius relation (right) predicted by our model for a fixed value of the boosting parameter ($K=20$) and different values for disruption ($1\leq\eta\leq20$). The tracks have been calculated assuming typical parameters, that is, initial donor mass and period of $M_{2} = 0.8 \, \rm{M_{\odot}}$ and $P_{\rm orb} = 1$\,day, respectively, and a constant white dwarf mass of $M_{\rm WD} = 0.83 \, \rm{M_{\odot}}$. 
    The mass-radius relation derived from observations \citep{mcallisteretal19-1} is reasonably well reproduced independent of the value of $\eta$ (right) but the fit improves for $\eta=2-3$ according to a $\chi^2$ test (see text). Such a moderate disruption is also required to generate a detached phase as an explanation for the orbital period gap \citep[shaded region and dashed vertical lines in the left and middle panels]{kniggeetal11-1,schreiberetal24-1}. These values of $\eta$ also predict a period minimum similar to that derived from observations \citep[][dotted and dashed-dotted vertical lines]{kniggeetal11-1,mcallisteretal19-1}. 
    Such a mild disruption is very different to (almost) fully turning magnetic braking off as assumed in the standard scenario of CV evolution. By fitting the mass-radius relation (right panel) we predict mass transfer rates above the gap that, on average, exceed those measured from white dwarf temperatures \citep[][left panel]{palaetal22-1}.}
    \label{fig:all}
\end{figure*}

Equipped with the revised SBD magnetic braking prescription, we ran several \mesa~simulations, assuming a donor star initial mass of $M_{2} = 0.8$ \msun~and initial orbital period $P_{\rm orb} = 1.0$ d, exploring the boosting and disruption parameters in the range $K = [10,20,50,80,100]$ and $\eta = [1,2,3,4,10,20]$. 
The observables that define these two parameters are the mass-radius relation of CV donor stars \citep{kniggeetal11-1,mcallisteretal19-1}, the boundaries of the period gap \citep{kniggeetal11-1,schreiberetal24-1}, and the period minimum \citep{kniggeetal11-1,mcallisteretal19-1}. The boosting parameter $K$ determines how bloated CV donors are compared to main sequence stars of the same mass and is crucial for the upper and lower edge of the period gap. The disruption parameter $\eta$ is critical for the lower edge of the period gap and the period minimum. Evolutionary tracks and observational constraints are compared in Fig.\,\ref{fig:all}. 

We find that the upper boundary of the period gap is well reproduced assuming $K\,\sim20$ which is smaller than previously estimated \citep[e.g.][]{barrazaetal25, bellonietal24-1}. This result is related to the recalibration of magnetic braking we performed which increases the non-boosted angular momentum loss in the saturated regime by a factor of $\sim\,4$ (see Fig.\,\ref{fig:saturation}).
Concerning the disruption parameter, we find even more drastic changes. 
In previous works $\eta \, \gappr \, 30-100$ was required \citep{bellonietal24-1,barrazaetal25,kniggeetal11-1}. 
For the recalibrated saturated prescription based on \citet{mattetal15-1} we find that  
$\eta$ needs to be just $2-3$ to reproduce the boundaries of the period gap and period minimum (see left panel of Fig.\,\ref{fig:all}.  

The observed mass-radius relation is reasonably well reproduced without any disruption at all ($\eta=1$) but seems to improve for small values $\sim2-3$ (right panel of Fig.\,\ref{fig:all}). 
We computed the reduced $\chi^{2}$ values for models with $K=20$ and $\eta=[1,2,3,4,10,20]$ and obtained 
 $\chi^{2}_{\rm red}\approx[60, 18, 32, 43, 67, 77]$, respectively, which confirms that indeed $\eta=2-3$ provides the best fit. 
 By fitting the observed mass radius relation, our model predicts mass transfer rates exceeding those derived from white dwarf temperatures \citep{palaetal22-1}. 

In summary, a moderate disruption ($\eta=2-3$) is sufficient to reproduce the period gap and the period minimum and is also indicated by the observed mass radius relation.
This result might have far reaching consequences for our understanding of magnetic braking and dynamo theory. While disruption is still required to reproduce the period gap, a drastic change which was typically thought to provide clear indications of a change in the dynamo mechanism \citep{taam+spruit89-1}, is not required anymore.

\section{Concluding discussion}

We recalibrated the magnetic braking prescription proposed by \citet{mattetal15-1} which leads to stronger angular momentum loss through magnetic braking in the saturated regime. Using this prescription and further increasing magnetic braking by a factor of $20$ we have shown that, in contrast to what has been assumed for decades in CV evolution theory, a relatively moderate decrease of the efficiency in angular momentum loss through magnetic braking by a factor of $\sim2-3$ can fully explain the key features of the observed orbital period distribution and the observed mass-radius relation. While this might imply that we are getting closer towards a unified magnetic braking prescription, it remains still unclear why the change in angular momentum loss at the fully convective boundary is the opposite in single stars \citep{luetal24-1} and why angular momentum loss through magnetic braking is in general stronger in close binaries. In what follows we discuss possible explanations. 

The strongest constraints on the secular mass transfer rates in CVs comes from the measured radii of the donor stars \citep{kniggeetal11-1}. These measured radii are typically larger than the radii of main sequence stars as calculated with MESA. This has been interpreted as compelling evidence for efficient magnetic braking in close binaries which is driving the mass transfer. 
However, recently, it has been shown that the radii of low-mass stars are larger than predicted by standard models, especially in magnetically active stars \citep{macdonald+mullan24-1,parsonsetal18-1, brownetal22-1}. This radius increase, most likely magnetically induced, is typically around $10$ per cent but may reach more than $20$ per cent in particular cases \citep{macdonald+mullan24-1}. 
This might imply that assuming all the bloating results from mass transfer driven by angular momentum loss might lead to significantly overestimating magnetic braking. Taking this effect into account might reduce the tensions in the strength of magnetic braking in single stars and in binaries as well as between the predicted mass transfer rates and those measured from white dwarf temperatures (see Fig.\,\ref{fig:all}, left panel).  

Concerning magnetic braking at the fully convective boundary we note several processes might play a role but have not been in detail 
considered in CV evolution . 
First, the self-consistently calculated convective turn-over time peaks at the fully convective boundary (see Fig.\,\ref{fig:all}), and therefore a dependence of magnetic braking on this time scale should be considered. Second, if we indeed significantly overestimate magnetic braking as outlined above, maybe the convective kissing instability, previously discussed by \citep{larsen+macdonald25-1}, could operate in CVs. 
Finally, the observed stalling in single stars is typically explained by internal angular momentum exchange between the convective envelope and the radiative core \citep[e.g.][]{curtisetal19-1,curtisetal20-1}.
 
In closing, we note that full validation for suggested magnetic braking prescriptions requires performing population synthesis which we plan to perform in the future also 
considering internal angular momentum redistribution and up-dated donor star radii in future models and plan to present the result in future publications. 

\begin{acknowledgements}
JBJ, MRS, and ADS thank for support from ero-STEP (DFG research unit, grant Schw536/37-1).      
\end{acknowledgements}

\bibliographystyle{aa}
\bibliography{bibs}

\end{document}